\documentclass[pre,reprint]{revtex4-1}

\usepackage{graphicx,amsmath,amssymb,color,bm}
\usepackage{natbib}

\newcommand{\bi}{\begin{itemize}}
\newcommand{\ei}{\end{itemize}}

\newcommand{\gdot}{\dot{\gamma}}

\newcommand{\be}{\begin{equation}}
\newcommand{\ee}{\end{equation}}
\newcommand{\beqna}{\begin{eqnarray}}
\newcommand{\eeqna}{\end{eqnarray}}

\newcommand{\tw}{t_w}

\newcommand{\sigmay}{\sigma_y}

\begin{document}

\title{Shear Banding of Soft Glassy Materials in Large Amplitude Oscillatory Shear}

\author{Rangarajan Radhakrishnan}%

\author{Suzanne M. Fielding}%
\email[Corresponding author: ]{suzanne.fielding@durham.ac.uk}
\affiliation{%
  Department of Physics, Durham University, Science Laboratories,
  South Road, Durham DH1 3LE, UK} \date{\today}

\begin{abstract}
We study shear banding in soft glassy materials subject to a large amplitude oscillatory shear flow (LAOS). By numerical simulations of the widely used soft glassy rheology model, supplemented by more general physical arguments, we demonstrate strong banding over an extensive range of amplitudes and frequencies of the imposed shear rate $\gdot(t)=\gdot_0\cos(\omega t)$, even in materials that do not permit banding as their steady state response to a steadily imposed shear flow $\gdot=\gdot_0=$const. Highly counterintuitively, banding persists in LAOS even in the limit of zero frequency $\omega\to 0$, where one might {\it a priori} have expected a homogeneous flow response in a material that does not display banding under conditions of steadily imposed shear.  We explain this finding in terms of an alternating competition within each cycle between glassy aging and flow rejuvenation.  Our predictions have far-reaching implications for the flow behavior of aging yield stress fluids, suggesting a generic expectation of shear banding in flows of even arbitrarily slow time variation.
\end{abstract}


\maketitle

Soft glassy materials (SGMs)~\cite{sollich1997,Fielding2014} such as emulsions~\cite{Becu:2006}, foams~\cite{Rouyer2008}, colloids~\cite{Mason1995,Knaebel2000}, microgels~\cite{Cloitre2000}, and star polymers~\cite{Rogers2010} share several notable features in their rheological (deformation and flow) properties.  Particularly striking is the phenomenon of aging~\cite{Fielding1999}, in which a material's rheological response becomes progressively more solidlike as a function of the sample age $\tw$, defined as the time elapsed after the sample was freshly prepared at time $t=0$, before a test deformation is applied at $t=\tw$.  A sustained applied shear flow can, however, halt aging and rejuvenate the sample to a steady state of effective age set by the inverse flow rate $1/\gdot$.  The steady state flow curve of shear stress $\sigma$ as a function of shear rate $\gdot$ then typically has a yield stress $\sigmay$ in the limit $\gdot\to 0$.  Such features have been attributed to the generic presence in soft glasses of structural disorder ({\it e.g.,} in a disordered packing of emulsion droplets) and metastability ({\it e.g.,} in the large energy barriers involved in stretching soap films, which impede droplet rearrangements).

In the experimental literature, SGMs are often also referred to as yield stress fluids (YSFs). Two distinct categories of YSFs have been identified: ``simple'' and ``viscosity bifurcating''~\cite{Moller2009,Fielding2014}.  Under a sustained applied shear flow, viscosity bifurcating YSFs~\cite{Ragouilliaux2007,Moller2009,Fall2010,Martin2012} exhibit a phenomenon known as shear banding, in which their steady state flow field comprises macroscopic bands of differing viscosities, with layer normals in the flow-gradient direction. This effect is thought to stem from a non-monotonicity in the underlying constitutive curve of shear stress as a function of shear rate (for initially homogeneous flow states).  In contrast, simple YSFs~\cite{Coussot2009,Ovarlez2010,Ovarlez2013} have a monotonic constitutive curve, which precludes banding in their steady state response to a sustained applied flow~\footnote{at least in the absence of concentration coupling}.

Despite this, simple YSFs often do display shear banding~\cite{Divoux2010,Divoux2011a,Divoux2011b,Moorcroft2011,Moorcroft2013a} during the time-dependent, transient process whereby a steady flow is established out of an initial rest state, following the switch-on of a constant shear rate $\gdot$ in a previously undeformed sample.  This banding is likewise transient, persisting only as long as it takes to establish a steady homogeneous flow state, consistent with the constitutive curve of simple YSFs being monotonic.

An important question of fundamental principle, therefore, is whether an imposed flow that has a {\em sustained} time dependence $\gdot(t)$ can give rise to correspondingly sustained shear banding, even in simple YSFs that lack banding as their steady state response to a time-independent flow, $\gdot={\rm const}$. Put simply, can heterogeneous flow arise simply as a consequence of the time dependence of an imposed deformation? Here we address this question by studying the exemplary protocol of oscillatory imposed shear, $\gdot(t)=\gdot_0\cos(\omega t)$, with supporting numerical data for other imposed waveforms~\cite{SM}, and by physical arguments suggesting that our results should indeed apply to time-varying flows more generally.  We show that banding is a key part of a material's flow response across a wide range of imposed amplitude and frequency $\gdot_0,\omega$.  Crucially, and counterintuitively, it persists even in the low-frequency limit $\omega\to 0$, due to a repeated competition within each cycle between aging and flow-rejuvenation, even though the true $\omega=0$ case of steady shear $\gdot=\gdot_0=$const precludes banding in simple YSFs. This has far-reaching implications for the flow behavior of aging glassy materials, suggesting a generic expectation of banding even in flows of arbitrarily slow time dependence.

The protocol of large amplitude oscillatory shear (LAOS)~\cite{Hyun2011} considered here is the focus of intense current interest in the rheology community, for its use in ``fingerprinting'' complex fluids via tests in which the strain rate amplitude $\gdot_0$ (or strain amplitude $\gamma_0=\gdot_0/\omega$) and frequency $\omega$ are separately varied. At high frequencies a material's elastic response is probed, while at low frequencies more viscous response might be expected (although we return below to reappraise that expectation for aging materials).  Large imposed amplitudes probe nonlinear response, with linear viscoelastic response recovered for small amplitudes.  

In YSFs, LAOS has been widely studied experimentally~\cite{Yoshimura1987,Knaebel2000,Viasnoff2003,Rouyer2008,Ewoldt2010,Renou2010,Guo2011,VanderVaart2013,Koumakis2013,Poulos2013,Poulos2015} and theoretically~\cite{Yoshimura1987,Viasnoff2003,Ewoldt2010,Rogers2011,Rogers2012,Koumakis2013,Mendes2013,Blackwell2014,Sollich1998,Rouyer2008}.  Few experiments have directly imaged the flow field, although strain localization has been reported in foam~\cite{Rouyer2008} and concentrated suspensions~\cite{Guo2011}.  All theoretical studies of which we are aware have simply assumed the flow to remain homogeneous, neglecting the possibility of banding. An important contribution of this Letter is to suggest that aging YSFs will generically exhibit shear banding in LAOS, and that this has a major influence on the measured bulk rheological signals.  This in turn suggests that attempts to rheologically fingerprint a material without taking banding into account should be treated with considerable caution.

We perform our study using the soft glassy rheology (SGR) model~\cite{sollich1997,Sollich1998}.  As described in Ref.~\cite{SM}, this considers an ensemble of elements, each corresponding to a local mesoscopic region of SGM.  Under an imposed deformation, each element experiences a buildup of local elastic stress, intermittently released by plastic relaxation events. These are modeled as hopping of the elements over strain-modulated energy barriers, governed by a noise temperature $x$, and with a microscopic attempt time $\tau_0=1$ (in our units).  Upon yielding, any element resets its local stress to zero and selects its new energy barrier at random from a distribution $\rho(E)=\exp(-E/x_g)$.  This confers a broad spectrum of yielding times $P(\tau)$ and results in a glass phase for $x<x_g=1$ (in our units), with rheological aging: in the absence of flow the typical relaxation time scale increases linearly with the system's age $\tw$. A sustained flow, however, rejuvenates the sample to an effective age set by the inverse flow rate $1/\gdot$. The constitutive curve $\sigma(\gdot)$ has a yield stress $\sigmay(x)$, beyond which it rises monotonically: this gives simple YSF behavior, precluding steady state banding.

In its original form~\cite{sollich1997,Sollich1998}, the SGR model only allows for spatially uniform flows and cannot address shear banding. It was then adapted to allow for heterogeneous flows in Ref.~\cite{Fielding2008}. (That study also modified the model to have a nonmonotonic constitutive curve, giving viscosity bifurcating YSF behavior. We remove that modification, to keep simple YSF behavior.) We perform waiting-time Monte Carlo~\cite{Voter2007,Fielding2008,Moorcroft2011,Moorcroft2013a} simulations of this spatially aware model, taking typically $m=100$ SGR elements on each of $n=25$ streamlines arranged in the flow-gradient direction $y=0\ldots L$, where $L$ is the gap size. (We have checked for convergence on increasing $m,n$.)  Slight stress diffusivity is needed to capture the slightly diffuse nature of any interface between shear bands~\cite{Lu1999,Olmsted2000a}, and is incorporated by transferring a fraction $w$ of the stress of any freshly yielded element to adjacent streamlines.  

Each run is initialized with a distribution of trap depths $P(E,t=0)=\rho(E)$, and all local stresses equal to zero, to model a freshly prepared sample.  We then allow it to age undisturbed for a time $\tw$ before imposing an oscillatory strain $\gamma(t)=\gamma_0 \sin[\omega(t-t_w)]$ for times $t>\tw$.  Once many ($N>50$) cycles have been executed we report the flow response.  All results shown are for a noise temperature $x=0.3$, but we have checked that all reported phenomena hold for $x=0.1$-$0.9$ in the glass phase $x<1$.

\begin{figure}[!t]
\includegraphics[width=3.5in]{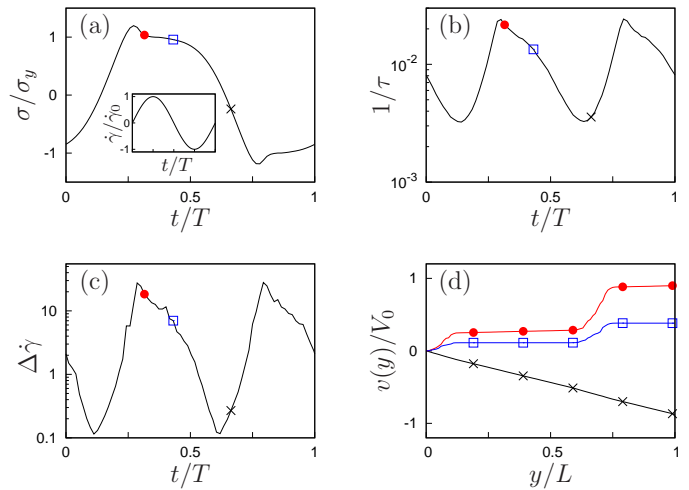}
\caption{Response of the SGR model to an imposed LAOS deformation with $\gamma_0=1.59$ and $\omega=0.01$.  Signals (shifted such that $\dot{\gamma}=0$ at $t=0$) show (a) shear stress and $\dot{\gamma}/\dot{\gamma}_0$ (inset), (b) inverse effective sample age, and (c) degree of banding over a cycle period $T=2\pi/\omega$.  Flow profiles are shown in (d) for the times indicated by the corresponding symbols in (a)-(c).  Noise temperature $x=0.3$, initial sample age $\tw=10$, cycle number $N=50$. $m=100$, $n=100$, $w=0.1$.} \label{fig:s_cyc}
\end{figure}

A typical cycle is shown in Fig.~\ref{fig:s_cyc}. Consider the first half of this, during which the strain rate is positive and the sample is straining forwardly.  At early times, when the strain rate has only just switched from negative to positive, the shear is weak and the sample is old and aging: the inverse effective sample age $1/\tau$~\footnote{$1/\tau=\sum_{i=1}^{i=n} \sum_{j=1}^{j=m}\exp [ - (E_{ij} - k l_{ij}^2)/x]/(m n)$, where $E_{ij}$ and $l_{ij}$ are energy trap depth and strain corresponding to an SGR element $ij$.  $k=1$ in our units and is assumed to be the same for all elements.} [Fig.~\ref{fig:s_cyc}(b)] is small and decreasing. Its rheological response is accordingly rather elastic, and the stress increases with the accumulating strain [Fig.~\ref{fig:s_cyc}(a)].  This shearing has the effect of then rejuvenating the sample: $1/\tau$ increases and attains a maximum, the stress displays an overshoot, and the sample yields into a flowing regime where the stress is relatively constant.  The same sequence then repeats (with appropriate sign changes) in the negative strain rate half of the cycle.

Closely associated with the rejuvenation, stress overshoot and yielding in each half cycle is the formation of shear bands.  At any time $t$ we quantify the degree of banding as the spatial variance in the shear rate across the streamlines $i=1..n$, giving $\Delta\gdot(t)=\sqrt{\langle \gdot_i^2\rangle-\langle\gdot_i\rangle^2}/\gdot_0$~\footnote{Recall that $\gdot_0$ is the amplitude of the imposed strain rate maximized over the cycle. Normalizing by this rather than the instantaneous $\gdot(t)$ therefore tends to give a conservative estimate of the degree of banding}. As seen in Fig.~\ref{fig:s_cyc}(c), the degree of banding increases sharply as the stress overshoot is approached and peaks shortly afterwards. The velocity profiles $v(y)=\int_{0}^{y} \dot{\gamma}(y') dy'$ then deviate strongly from linearity [Fig.~\ref{fig:s_cyc}(d)], consistent with banded $\gdot(y)$ profiles.

So far, we have discussed the response over one particular cycle: the $N=50$th after the inception of the flow. For an aging system, however, an important question is whether this intracycle response is invariant from cycle to cycle, $t\to t+2\pi/\omega$, for large enough $N$. Indeed, in the absence of shear banding, indefinite cycle-to-cycle aging is expected for all strain amplitudes~\cite{Viasnoff2003}.  However, in any regime where significant banding arises, we have verified that after typically $N=50 $ cycles there is no noticeable further change in the stress $\sigma (t)$ or degree of banding $\Delta \gdot (t)$ from cycle to cycle $t\to t+2\pi/\omega$~\cite{SM2}.  Furthermore, the flow state then has no memory of the sample age $\tw$ before shearing commenced, with runs for initially old ($\tw=10^6$) and young ($\tw=10$) samples having converged (data not shown).


%
\begin{figure}[!t]
\includegraphics[width=3.5in]{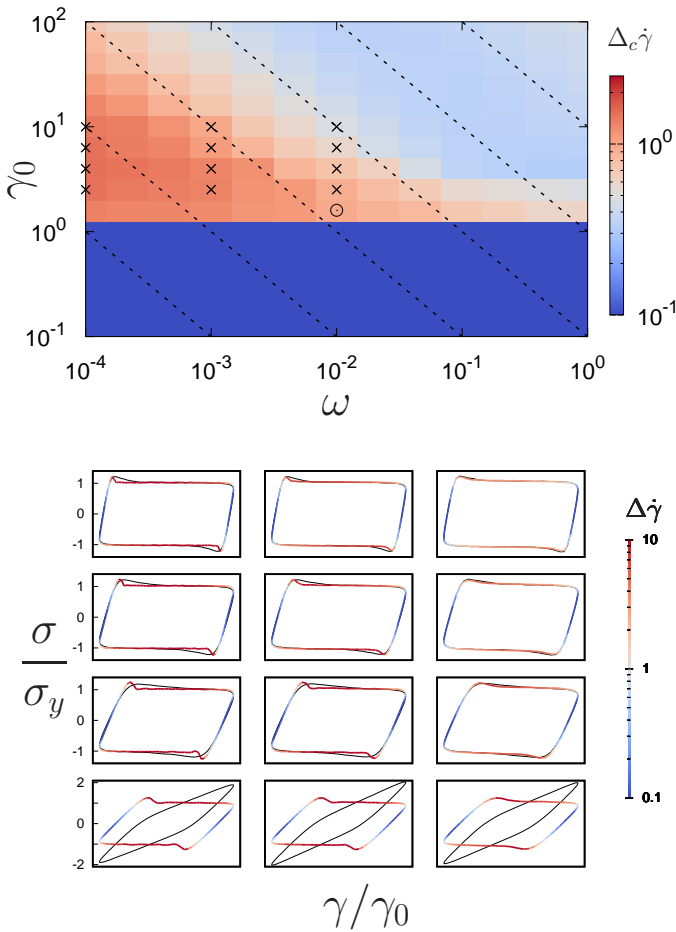}
\caption{Top: Dynamic phase diagram showing as a color map the cycle-averaged degree of banding in the SGR model in LAOS. Dashed lines show constant $\dot{\gamma}_0$. Bottom: Elastic Lissajous-Bowditch curves for the homogeneous (dashed lines), and heterogeneous (solid lines) SGR model, with the instantaneous degree of banding $\Delta \dot{\gamma}$ indicated by the color scale, for the grid of $\gamma_0,\omega$ values shown by $\times$ in the top panel. (The circle in the top panel shows $\gamma_0,\omega$ values of the run in Fig.~\ref{fig:s_cyc}.) Noise temperature $x=0.3$, initial sample age $\tw=10.0$.  $m=100$, $n=25$, $w=0.05$.} \label{fig:contour}
\end{figure}

In Fig.~\ref{fig:s_cyc} we discussed in detail the response of the SGR model to a single LAOS experiment for one particular set of imposed amplitude and frequency values $(\gamma_0,\omega)$. To summarize more broadly the regimes of $(\gamma_0,\omega)$ in which banding arises, we show in Fig.~\ref{fig:contour} (top) a dynamic phase diagram where each coordinate pair $(\gamma_0,\omega)$ corresponds to a LAOS experiment with those given $(\gamma_0,\omega)$.  Represented by the colored block at each $(\gamma_0,\omega)$ is the degree of banding that arises in that particular LAOS experiment, averaged over a cycle: $\Delta_c \gdot=\langle\Delta\gdot(t)\rangle_T$~\footnote{Indeed, to reduce noise we further average $\Delta_c$ over the $N=50$-$100$th cycles.  Recall that in any regime where banding is significant, $\Delta_c\gdot$ becomes cycle-to-cycle invariant after 50 cycles}.  A value $\Delta_c\gdot > 0.5$ (red region) corresponds to strongly visually apparent banding in the flow profiles.  As can be seen, significant banding arises in a large region of the $(\gamma_0,\omega)$ plane: (roughly) for $\gamma_0> 1$ and $\gdot_0=\gamma_0\omega < 0.1$. (The contour $\gamma_0\omega=0.1$ is shown by the fourth dashed line from the left in Fig.~\ref{fig:contour}, top.)  Elsewhere the flow remains homogeneous. The transition from homogeneous to banded flow around $\gamma_0=1.0$ looks sharp in Fig.~\ref{fig:contour} (top), due to the discretization used, but in fact occurs over a typical strain scale $\Delta\gamma_0\approx 0.1$. This will be discussed in more detail in a future paper.

Further insight into the physical processes within each cycle can be gained by parametrically plotting the stress as a function of strain, to give the so-called elastic Lissajous-Bowditch (ELB) curve.  In this representation, a linear elastic solid would give a straight line through the origin, a viscous liquid an ellipse.  At low strain amplitudes, the ELB curves of the SGR model indeed indicate rather elastic response (not shown). In contrast, for strain amplitudes $\gamma_0>1$ the alternating competition within each cycle between aging and rejuvenation, and between elastic and viscous response, gives highly nonlinear ELB curves as observed in soft glasses~\cite{Renou2010,Rogers2011,Poulos2015}.

This is seen in the solid curves in the bottom panel of Fig.~\ref{fig:contour}, which shows a grid of ELB curves (a ``Pipkin diagram'') corresponding to the grid of $(\gamma_0,\omega)$ values indicated by crosses in the top panel. (The dashed curves in the same figure will be discussed below.) Over the course of a LAOS cycle any ELB curve is explored once in the clockwise direction, with the bottom-left to top-right sector corresponding to the positive strain rate half of the cycle.  A sequence of physical processes~\cite{Rogers2011} can be identified as follows.  At the bottom left of the ELB curve the strain rate has just switched from negative to positive and the sample is being sheared only weakly.  As a result it is aging and shows rather elastic response, with the stress initially increasing linearly with strain.  This shearing then has the effect of rejuvenating the sample, which eventually yields: the stress goes through an overshoot then declines to a flowing regime where it barely changes with strain. The color scale in each ELB curve shows the degree of banding at that point in the cycle. As can be seen, the onset of banding is closely associated with the stress overshoot.

For an ergodic viscoelastic material with a fixed characteristic relaxation time $\tau$, a progression from elasticlike ($\omega\tau\gg 1$) to viscouslike ($\omega\tau\ll 1$) response is expected in a sequence of LAOS experiments repeated at progressively lower frequency $\omega$ (albeit also with nonlinear effects for $\gamma_0>1$).  In particular, in the limit $\omega\to 0$ we expect to recover a regime in which the fluid quasistatically explores its viscous steady state flow behavior as the strain rate sweeps slowly up and down in each cycle.  A Lissajous-Bowditch curve plotted in the viscous representation $\sigma(\gdot)$ (VLB) should then simply correspond to the steady state flow curve.  For any material with a monotonic underlying constitutive curve, shear banding would be impossible in this quasistatic limit~\cite{Adams2009,Carter2015}.

\begin{figure}[!t]
\includegraphics[width=3.5in]{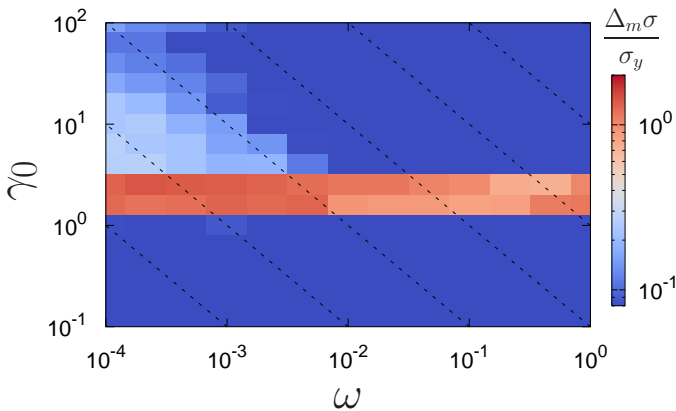}
\caption{Maximum discrepancy between stress predicted by a calculation that assumes homogeneous flow and by the full heterogeneous calculation. Parameter values as in Fig.~\ref{fig:contour}.}
\label{fig:dev}
\end{figure}

Surprisingly, no such progression is evident with decreasing frequency leftwards along any row of the Pipkin grid in Fig.~\ref{fig:contour} (bottom): even at the lowest frequencies, the sample still shows strongly elastic response in some part of the cycle, with the stress increasing linearly with strain.  The VLB curve $\sigma(\gdot)$ (not shown) never approaches the steady state flow curve, and instead always has markedly open loops. Strong shear banding still occurs, despite the underlying constitutive curve being monotonic.  This highly counterintuitive behavior is due to a basic, alternating competition within each cycle between aging (in the low shear phase) and rejuvenation, yielding and banding (in the high shear phase). This finding has far-reaching implications for the flow of aging soft glasses, suggesting a generic expectation of shear banding even in protocols of arbitrarily slow time variation~\cite{SM}.  Put simply: an aging material has no fixed characteristic relaxation time scale $\tau$ against which to compare the frequency $\omega$ of the imposed oscillation. The highly nonlinear, nonequilibrium phenomenon of shear banding can therefore persist even to arbitrarily low frequencies.

Most theoretical studies of LAOS to date have assumed homogeneous flow, neglecting the possibility of banding.  Our results in Fig.~\ref{fig:contour} (bottom) show this assumption can be very misleading: in each panel the solid line shows the ELB curve in a calculation that allows for banding and the dashed line shows the same curve in a calculation that assumes homogeneous flow.  Banding clearly causes a strong discrepancy between these, particularly for strain amplitudes only just in the nonlinear regime, which are most commonly studied experimentally.  To explore this further, in Fig.~\ref{fig:dev} we map the regions of the $\gamma_0,\omega$ plane in which this discrepancy is most pronounced, by showing as a color scale at each $(\gamma_0,\omega)$ the maximum difference in stress $\Delta_m \sigma$ between the homogeneous and heterogeneous calculations~\footnote{For numerical convenience, the difference is measured over a time interval $T/10$ following the peak in the stress signal for the heterogeneous flow, which is indeed when any difference between these two signals is most pronounced}.  An important additional message of this work is, therefore, to council considerable caution in seeking to fingerprint complex fluids by calculations that assume homogeneous flow.

In the simpler protocol of shear startup, an initially undeformed sample is subject to the switch-on of shear of some rate $\gdot$.  Simple YSFs often shear band transiently in this protocol, around the time of the overshoot in the start-up signal of stress as a function of accumulated strain, with the effect being more pronounced for initially older samples~\cite{Divoux2010,Divoux2011a,Moorcroft2011,Moorcroft2013a}.  In view of this, it is tempting loosely and intuitively to interpret LAOS as a repeated alternation of forward then reverse start-up runs, with banding triggered by the stress overshoot in each half cycle.  This connection is, however, far from clear: there is no reason, {\it a priori}, why the simpler protocol of shear startup should inform the much more complicated one of LAOS. Whether it is possible to establish any rigorous correspondence between these remains open.

To summarize, we have demonstrated that soft glasses may exhibit shear banding when subject to an applied shear flow with a sustained time-dependence, using large amplitude oscillatory shear as an illustrative example (with supplemental data for other waveforms in~\cite{SM}).  Counterintuitively, strong banding persists even in the limit of arbitrarily slowly time-varying flows. This is true even in materials that do not show banding as their steady state response to shear of a constant rate, and, therefore, in which one would {\it a priori} have expected a quasistatic, homogeneous flow response in the low-frequency limit.  We have shown that this can be understood in terms of an alternating competition between aging and flow rejuvenation within each cycle.  This finding has potentially far reaching implications for the flow of aging soft glasses, suggesting a generic expectation of banding even in flows of arbitrarily slow time variation.

The research leading to these results has received funding from the European Research Council under the EU's 7th Framework Programme (FP7/2007-2013) / ERC Grant No. 279365. The authors thank K. A. Carter for discussions, P. Sollich for data to check our code in homogeneous flow, and P. Sollich and M. E. Cates for a critical reading of the manuscript.

\newpage
\clearpage
\renewcommand{\figurename}{SM.FIG.}
\setcounter{figure}{0}  
\section*{Supplemental material for Shear Banding of Soft Glassy Materials in Large Amplitude Oscillatory Shear}

\subsection{Soft Glassy Rheology Model}

The soft glassy rheology (SGR) model~\cite{sollich1997,Sollich1998} considers an ensemble of elements undergoing activated hopping in an energy landscape of traps.  Each element is taken to correspond to a local mesoscopic region of a soft glassy material, and is assigned continuum variables of local shear strain $l$ and stress $kl$, which describe elastic deformation relative to a locally undeformed equilibrium.  Between hops, each element affinely follows the macroscopic strain field, $\dot{l}=\gdot$. A local yielding event is then identified as the hopping of an element out of one trap and into another.  Hops are modelled as being dynamically activated, with an element in a trap of depth $E$ and with local shear strain $l$ having probability per unit time of yielding $\tau^{-1}(E,l)= \tau_0^{-1} \exp \left(-(E-\frac{1}{2}kl^2)/x\right)$.  The parameter $x$ is an effective mean field noise temperature, intended to model coupling with other yielding events elsewhere in the sample. After hopping, the element selects a new trap depth randomly from a prior distribution $\rho(E)\sim\exp(-E/x_g)$, and to reset its strain $l$ to zero.

With these dynamics, the probability $P(E,l,t)$ of an element to be in a trap of depth $E$ with local shear strain $l$ evolves according to
\be
\label{eqn:master}
\dot{P}(E,l,t)+\gdot\frac{\partial P}{\partial l} = -\frac{1}{\tau(E,l)}P+Y(t)\rho(E)\delta(l).
\ee
The convected derivative on the left hand side describes affine
loading of each element by shear. The first and second terms on the
right hand side describe hops out of and into traps respectively, with
an ensemble average hopping rate
\be
Y(t)=\int dE \int dl \frac{1}{\tau(E,l)}P(E,l,t).
\ee
The macroscopic stress of the sample as a whole is defined as the
average over the local elemental stresses:
\be
\sigma(t)=\int dE \int dl\; kl P(E,l,t).
\ee

So far we have described the SGR model in its original, spatially
homogeneous form~\cite{sollich1997,Sollich1998}. The model can be extended~\cite{Fielding2008,Moorcroft2013a} to account
for shear banding in the flow gradient dimension $y$ by discretizing
the $y$ coordinate into $i=1 \cdots n$ streamlines of equal spacing
$L_y/n$, for a sample thickness $L_y$.  Each streamline is assigned
its own ensemble of $j=1\cdots m$ SGR elements, with a streamline
shear stress $\sigma_i=(k/m)\sum_j l_{ij}$. In creeping flow, the
shear stress must remain uniform across all streamlines at all times,
$\sigma_i(t)=\sigma(t)$.  Supposing a hop occurs at element $ij$ when
its local strain is $l=\ell$, reducing the stress on that streamline,
and so potentially violating force balance. Force balance can then be
restored by updating all elements on the same streamline $i$ as $l\to
l + \ell/m$, to ensure uniform stress across streamlines, then
updating all elements on all streamlines as $l \to l - \ell/mn$ to
ensure a properly reduced global stress.  A small stress
diffusivity between neighbouring streamlines is incorporated
by further adjusting the strain of three randomly chosen elements on
each adjacent streamline $i\pm 1$ by $\ell w(-1,+2,-1)$.

To simulate the model just described, we use a waiting time Monte Carlo algorithm~\cite{Voter2007,Fielding2008}. Given a hop rate $r_{ij}=\tau^{-1}(E_{ij},l_{ij})=\tau_0^{-1}\exp\left[-(E_{ij}-\tfrac{1}{2}kl_{ij}^2)/x\right]$ for the $j$th element on the $i$th streamline, the next element to yield is selected stochastically according to the weighted probability $P_{ij}=r_{ij}/\sum_{ij}r_{ij}$, with $P_{ij}$ being the probability that the next element to yield is the $j$th particle in the $i$th streamline. The time interval to the next hop is likewise selected stochastically as $dt=-\ln(s)/\sum_{ij}r_{ij}$, where $s$ is a random number selected from a uniform distribution between $0$ and $1$. All the results reported are converged with respect to increasing number of streamlines $n$ and elements per streamline $m$.

\subsection{Robustness of  banding scenario to the imposed waveform}

\begin{figure}[!ht]
\includegraphics[width=3.5in]{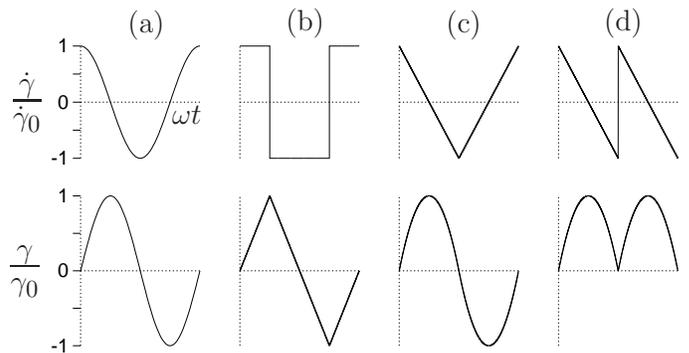}
\caption{Sketch of the strain-rate waveforms considered: a) oscillatory (as studied in detail in the main text), b) square wave c) triangle wave, and d) sawtooth. Top row: strain rate profiles over one cycle normalized by the maximum strain rates $\dot{\gamma}_0$ which are (a) $\omega\gamma_0 $, (b) $4 \omega\gamma_0 $, (c) $8 \omega\gamma_0$, and (d) $8 \omega\gamma_0$. Bottom row: strain profiles normalized by the maximum imposed strain $\gamma_0$.} \label{fig:imposed_flow}
\end{figure}

In the main text we demonstrated shear banding to be an important feature of the response of a soft glassy material in large amplitude oscillatory shear (LAOS), which has an imposed strain rate waveform $\gdot(t)=\gdot_0\cos(\omega t)$, for a broad range of values of amplitude $\gdot_0$ and frequency $\omega$. (We set the start time of the oscillation to zero in this supplementary material for simplicity of notation. In the main text, we noted that the absolute start time $\tw$ anyway becomes irrelevant after many cycles have been executed.) We also provided arguments suggesting that this scenario should hold more generically for flows such as this with a sustained time-dependence and noted that this should remain true even in the limit in which that time variation is arbitrarily slow, $\omega\to 0$: a counterintuitive result, given that the monotonic underlying constitutive curve of simple yield stress fluids precludes banding in the true zero frequency case of a steadily imposed shear flow, $\omega=0$.  

\begin{figure*}[!ht]
\includegraphics[width=5.5in]{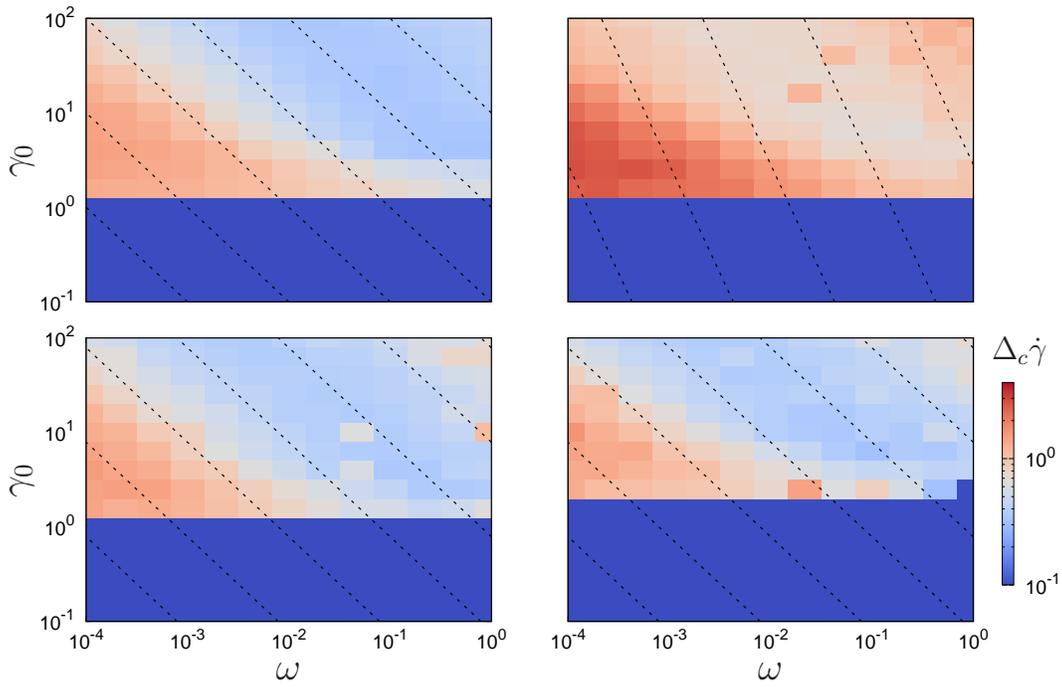}
\caption{Dynamic phase diagram showing as a color map the
  cycle-averaged degree of banding in the SGR model in a) LAOS, b)
  square wave strain rate, c) triangle wave strain rate and d)
  sawtooth wave strain rate. Dashed
  lines show lines of constant $\dot{\gamma}_0$.  Noise temperature
  $x=0.3$, initial sample age $\tw=10.0$, $m=100$, $n=25$, $w=0.05$.}
\label{fig:SB_response}
\end{figure*}

Here we support this claim to generality by presenting numerical
results for other imposed strain rate waveforms: square wave, triangle
wave, and sawtooth wave, as sketched in
SM.Fig.~\ref{fig:imposed_flow}b-d), alongside the original LAOS protocol
of the main text (panel a).  State diagrams showing the degree of
shear banding as a function of the amplitude and frequency of the
imposed strain rate signal are shown in
SM.Fig.~\ref{fig:SB_response}b-d), alongside the original LAOS results of
the main text (panel a). As can be seen, the same scenario holds in
all four protocols.

\subsection{Cycle-to-cycle invariance}

\begin{figure}[!h]
	\centering
\includegraphics[width=3.5in]{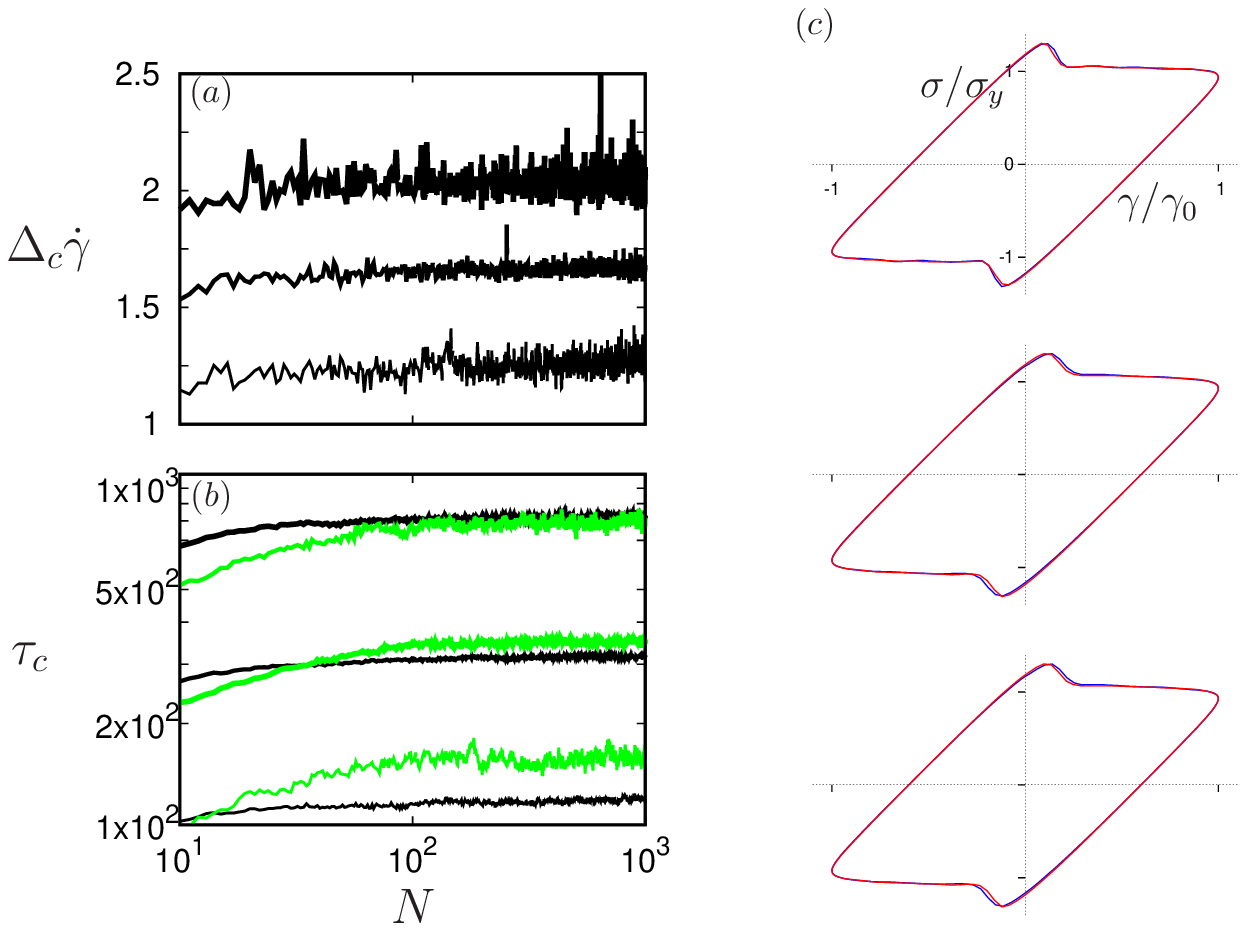}
\caption{A study of cycle-to-cycle invariance versus cycle-to-cycle
  aging in the SGR model in large oscillatory shear flow at a noise
  temperature $x=0.3$. Imposed strain amplitude $\gamma_0=1.59$ (which
  is the lowest non-linear value for which we find appreciable shear
  banding) for increasing cycle frequencies
  $\omega=10^{-3},10^{-2.5},10^{-2}$ (top to bottom in each panel).
  {\bf (a)} Cycle-averaged degree of banding $\Delta_c \dot{\gamma}$
  versus cycle number $N$. {\bf (b)} Cycle-averaged sample age
  $\tau_c$ versus cycle number. Black lines: heterogeneous model with
  $m=100$, $n=100$ and $w=0.05$. Green lines: homogeneous model with
  $m=1000$, and $n=1$. {\bf(c)} Comparison of elastic Lissajous
  Bowditch curves for cycle number $N=50$ (blue lines) and $N=1000$
  (red lines, which are essentially indistinguishable from the blue
  ones).}
\label{fig:ageing}
\end{figure}

In the main text, we showed numerical results typically for cycle
number $N=50$ to $N=100$ after the inception of the flow, and stated
that the relevant rheological quantities become essentially invariant
from cycle-to-cycle $t\to t + 2\pi/\omega$ after (typically) 50 cycles
have been executed. Here we present numerical data supporting that
claim.  SM.Figs.~\ref{fig:ageing}a) and c) show that the degree of
banding $\Delta_c\dot{\gamma}$ and stress signals are essentially
constant between cycles $N=50$ and $N=1000$.  Some slow upward drift
in the degree of banding is perhaps evident for the larger frequency
runs in panel a), although it is within its own noise.  More
importantly, there is certainly no evidence for any gradual {\em
  decrease} in the degree of banding with cycle number, which might
have undermined the scenario uncovered in this work. Of course the
true limit $N\to \infty$ cannot be addressed by numerical simulation.
However in experimental practice the number of cycles executed at the
low frequencies of interest here would in practice be very highly
unlikely to exceed $O(10)-O(100)$.

Some very slow increase is however apparent in the sample age $\tau_c$
as a function of cycle number. See SM.Fig.~\ref{fig:ageing}b), which
separately shows results for the full shear banding simulations (black
lines) and for simulations in which the flow is artificially
constrained to remain homogeneous (green lines). Indeed, previous
studies -- which assumed homogeneous flow -- argued that indefinite
cycle-to-cycle aging must occur for any imposed $\gamma_0$ ~\cite{Viasnoff2003},
because energy traps of the SGR model with depths $E$ greater than a
cutoff energy $E_c=\gamma_0^2/2+x \ln(t)$ are not perturbed by imposed
strain, whereas elements trapped in shallower traps are rejuvenated by
it.  This is consistent with the (very slow) aging apparent in the
green lines. However, this appears mitigated in the spatially
aware simulations in which banding arises.

\bibliographystyle{apsrev4-1}

\end{document}